\documentclass[%
twocolumn,footinbib,
superscriptaddress,
floatfix,
amsmath,amssymb,
aps,
prl, 
longbibliography,
groupedaddress,
]{revtex4-2}
\usepackage{graphicx}
\usepackage{physics}
\usepackage{amsmath}
\usepackage{natbib}
\usepackage{soul}
\usepackage{xprintlen}
\usepackage{bbold}
\usepackage{color}
\usepackage{dcolumn}
\usepackage{xr}
\usepackage[colorlinks=true, pdfstartview=FitV, linkcolor=blue, citecolor=blue, urlcolor=blue]{hyperref} 
\usepackage{bm}
\usepackage{hyperref}

\begin{document}
	
	\preprint{APS/123-QED}
	
	\title{Second-order topological modes in two-dimensional continuous media}
	
	\author{Jan Ko\v{s}ata}
	
	\author{Oded Zilberberg}
	\affiliation{Institute for Theoretical Physics, ETH Z{\"u}rich, 8093 Z{\"u}rich, Switzerland}
	\date{\today}
	
	\begin{abstract}
		We present a symmetry-based scheme to create 0D second-order topological modes in continuous 2D systems. We show that a metamaterial with a \textit{p6m}-symmetric pattern exhibits two Dirac cones, which can be gapped in two distinct ways by deforming the pattern. Combining the deformations in a single system then emulates the 2D Jackiw-Rossi model of a topological vortex, where 0D in-gap bound modes are guaranteed to exist. We exemplify our approach with simple hexagonal, Kagome and honeycomb lattices. We furthermore formulate a  quantitative method to extract the topological properties from finite-element simulations, which facilitates further optimization of the bound mode characteristics. Our scheme enables the realization of second-order topology in a wide range of experimental systems.
	\end{abstract}
	\maketitle

	The development of topological insulators (TIs), while originating in electronic systems, has made a profound impact in the field of classical metamaterials. Their signature topological boundary modes have been successfully demonstrated in photonic~\cite{Li_2018, Ozawa_2019}, electrical~\cite{Ningyuan_2015, Hofmann_2019}, phononic~\cite{Nash_2015,  Mousavi_2015, Huber_2016}, acoustic~\cite{He_2016, Ma_2019}, atomic~\cite{Cooper_2019} and polaritonic~\cite{Milicevic_2015} systems. This remarkable universality stems from the origin of the boundary phenomena; these depend only on topological invariants derived from the bulk spectral bands, and not on the specific medium. Early realizations of topology in classical systems relied on engineering synthetic gauge fields in tight-binding (TB) models, obtained by coupling resonator modes in space~\cite{Kraus_2012,Hafezi_2013} and/or time~\cite{Rechtsman_2013} to create topological boundary modes. Later it was discovered that in patterned continuous media, topologically distinct phases can be formed purely by breaking the spatial symmetries of the pattern~\cite{Wu_Hu_2015}; this has opened the field of TIs to a wider range of platforms, and brought it closer to potential applications. 
	Furthermore, the topological phases of matter paradigm has been recently extended to higher-order TIs (HOTIs)~\cite{Kraus_2013,Benalcazar_2017a,Benalcazar_2017b,Petrides_2018, Zilberberg_2018, Benalcazar_2019, Calugaru_2019, Fukui_2019, Petrides_2020}. In a $d$-dimensional HOTI, topological modes of dimension $d-2$ and lower may appear. HOTIs have been experimentally demonstrated by emulating TB models in a variety of platforms~\cite{Imhof_2018, Noh_2018, Schindler_2018, Serra-Garcia_2018, Fan_2019, Xue_2019, Xie_2018, Zhang_2019, Zhang_2020, Zhou_2020}.
	
	The concepts of TIs and HOTIs are unified in the framework of ten topological classes, each characterized by time reversal, chiral and particle-hole symmetries~\cite{Chiu_2016}. For a $d$-dimensional system with a defect defined on a $\mathcal{D}$-dimensional surface, a topological invariant exists whose possible values depend only on the bulk symmetry class and $d-\mathcal{D}$~\cite{Teo_2010}. At such a defect, modes with dimension $d-1-\mathcal{D}$ (codimension $\mathcal{D} + 1$) are formed. TIs, with their topology defined purely in their bulk, correspond to $\mathcal{D} = 0$; for $\mathcal{D}>0$, the invariant generally includes integrals over curvatures combining real and momentum space, leading to the appearance of high-order topological modes~\cite{Teo_2010, Kraus_2013, Benalcazar_2017a, Benalcazar_2017b, Petrides_2018, Zilberberg_2018,Calugaru_2019, Fukui_2019,Petrides_2020} with topological boundary states of codimension $>1$. An instance of a HOTI is found in the $d=2$ Jackiw-Rossi Hamiltonian of a topological vortex, which can be implemented by deforming the TB model of graphene~\cite{Hou_2007}. Emulating this model has successfully demonstrated the creation of 0D modes at topological vortices in photonic~\cite{Gao_2019, Gao_2020,  Yang_2020} and elastic~\cite{Xiaoxiao_2021} devices. Compared with standard 0D modes formed at non-topological lattice defects, topological vortex modes offer several advantages~\cite{Gao_2020}.    (i) Their frequency is near mid-gap, resulting in better spatial confinement and quality factors. (ii) The number of modes formed is fixed by topology.  (iii) The modal area is scalable, as the modes form independently of the vortex size. These aspects make topological vortex modes promising candidates for the construction of single-mode semiconductor lasers.

\begin{figure}
		\centering
		\includegraphics[width=\columnwidth]{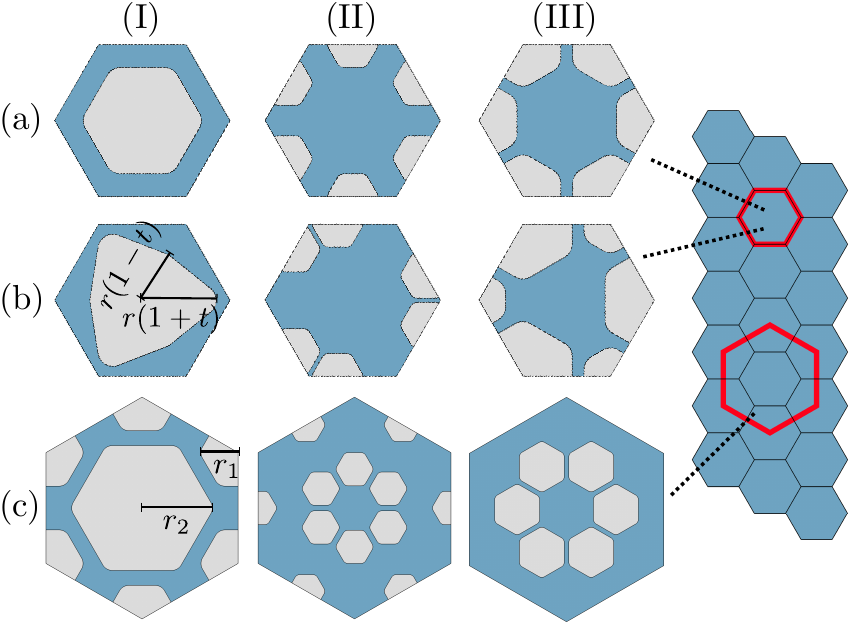}
		\caption{Metamaterial unit cells for the continuum realization of the Jackiw-Rossi model, cf.~Eq.~\eqref{eq:H}. The cells are patterned by two different linear media, e.g., a dielectric (blue) with air incisions (grey). The three columns show (I) the simple hexagonal lattice, (II) the Kagome lattice, and (III) the honeycomb lattice. The first row (a) corresponds to the starting unit cells, all of which form lattices with the \textit{p6m} space group. To obtain the anticommuting mass terms required in the Jackiw-Rossi model, we consider two symmetry-breaking perturbations; in row (b) breaking inversion symmetry [cf.~Eq.~\eqref{eq:M1}], and in row (c) breaking the translation symmetry of the primitive cell [cf.~Eq.~\eqref{eq:M2}]. This does not depend on the specific shapes of the incisions, as long as the respective symmetries are obeyed. The small and large red hexagons illustrate the unit cell before and after breaking translation symmetry. The labels $r, t, r_1, r_2$ in column (I) parameterize the incision shapes.}
		\label{fig:fig1}
	\end{figure}

	In this Letter, we show how to obtain the Jackiw-Rossi Hamiltonian in continuous 2D structures. Crucially, our approach does not require a recourse to a TB model or threading by magnetic fluxes. Instead, we utilize solely the breaking of spatial symmetries of a linear medium. We focus on the \textit{p6m} space group, picking three examples of metamaterial patterns: the simple hexagonal~[Fig.~\ref{fig:fig1}(I)(a)], Kagome~[Fig.~\ref{fig:fig1}(II)(a)], and honeycomb~[Fig.~\ref{fig:fig1}(III)(a)] lattices. Central to our work, the group \textit{p6m} has a 4D irreducible representation that manifests as two Dirac cones at the $K/K'$ points in reciprocal space. We consider two distinct perturbations of the primitive cell, namely the breaking of inversion and translation symmetries, and find their matrix representations in the 4D eigenspace~\cite{Saba_2020, Cano_2018}. The perturbation matrices are shown to anticommute and thus correspond to different gap-opening terms; these constitute the Jackiw-Rossi model. By spatially varying the perturbations in a single system, a topological vortex defect is formed with in-gap 0D  modes bound within. For concreteness, we assume a dielectric implementation of the scheme; we stress however that our analysis holds for any patterned 2D linear medium and is thus applicable to a wide range of systems.

	The 2D Jackiw-Rossi Hamiltonian reads
	\begin{equation} \label{eq:H}
	H = \boldsymbol{\gamma} \vdot \boldsymbol{k} + \boldsymbol{\Gamma} \vdot \boldsymbol{m}(\boldsymbol{r})\,,
	\end{equation}
	where $\boldsymbol{k} = (k_x, k_y)$ are quasimomenta, and $\boldsymbol{m}=(m_1,m_2)$ are mass terms that can change in space $\boldsymbol{r}=(x,y)$. The four $4 \times 4$ matrices, $\boldsymbol{\gamma} = (\gamma_1, \gamma_2)$ and $\boldsymbol{\Gamma} = (\Gamma_1, \Gamma_2)$, anticommute with one another.
	If either of the mass terms $m_1$ or $m_2$ is nonzero, the spectrum of $H$ is gapped around zero energy. We first consider the case where one mass term (e.g., $m_2$) vanishes and the other switches signs in space [e.g., $m_1 = \abs{m}\, \text{sgn}(x)$], thus defining a domain wall at $x=0$. The distinct band topology of the two bulk phases then defines a topological defect ($\mathcal{D}=0$) and a band inversion occurs across $x=0$. The domain wall can then be described by two copies of the Jackiw-Rebbi model~\cite{Jackiw_1976, Wu_Hu_2015}, and the system admits a $\mathbb{Z}_2$ invariant~\cite{Kane_2005b, Qi_2008}. Correspondingly, two counter-propagating 1D edge states appear at the domain wall. Depending on the eigenbasis of the gapping term, we thus obtain the states associated with either the quantum spin Hall effect (QSHE) or the quantum valley Hall effect (QVHE).
	
	Allowing instead both masses to vary over a space-dependent closed path defines a defect with $\mathcal{D} = 1$~\cite{Teo_2010}. The Hamiltonian is then characterized by a $\mathbb{Z}$ topological invariant associated with the winding number $n$ of $\boldsymbol{m}(\boldsymbol{r})$. Taking $m_1 + i m_2 = \abs{\boldsymbol{m}} e^{i \theta}$, the winding number is defined as
	\begin{equation} \label{eq:invariant}
	n = \frac{1}{2\pi} \oint_{S^1} d\theta\,,
	\end{equation}
	where $S^1$ denotes the closed path. At such a topological vortex, $n$ zero-dimensional bound modes are found. Systems with $\abs{n} > 0$ can be constructed within the TB model of graphene, where $m_1$ and $m_2$ are generated by spatial perturbations of the six-site unit cell~\cite{Hou_2007}. We also emphasize that this model is topologically equivalent to other proposed TB models for HOTIs~\cite{Fukui_2019} and the invariant definition involves integrating over both real and momentum space, such that the winding encircles a quantized 4D Dirac cone~\cite{Petrides_2020}.
	
	\textit{Symmetry analysis.---} We focus on the \textit{p6m} space group, exemplified by the three unperturbed structures shown in Fig.~\ref{fig:fig1}(a). All three are symmetric under six-fold rotation $C_{6}$, reflection $\sigma_{x}$, and lattice translation $T$.
	The operations that leave the structures invariant form a group $\mathcal{G}$ that is isomorphic to \textit{p6m}. The group $\mathcal{G}$ admits a 4D irreducible representation (irrep) derived from the little group of the $K/K'$ points~\cite{Bradley_2009}. This irrep manifests itself as the familiar pair of Dirac cones, appearing at the valleys $K$ and $K'$ in reciprocal space. Choosing a basis for this irrep defines a function $\rho(g)$, which maps each symmetry element $g \in \mathcal{G}$ to a $4\times 4$ representation matrix $\rho(g)$. For the three group generators, we obtain~\cite{supmat}
	\begin{gather}
	\rho(C_6) = \tau_1 \otimes R_6 \,, \quad \rho(\sigma_{x}) = \tau_1 \otimes \tau_3 \,, \nonumber\\
	\rho(T) = \text{diag}(w^2, w) \otimes \mathbb{1}_2\, ,
	\label{eq:operations}
	\end{gather}
	where $\tau_i$ are Pauli matrices, $R_6$ denotes a $2\times 2$ rotation matrix of angle $\pi/3$, and $w = \text{exp}\{2 \pi i / 3\}$. The basis is chosen such that each $\rho(g)$ is a direct product of a $2 \times 2$ matrix acting on the two valley degrees of freedom with another $2 \times 2$ matrix acting on the two degenerate bands within each valley.
	
	Opening band gaps in the degenerate cones to create topologically distinct bulks can be achieved in two different ways: breaking inversion or translation symmetry, as shown in Fig.~\ref{fig:fig1}(b) and \ref{fig:fig1}(c), respectively. Starting with the former, we break inversion symmetry while preserving $C_3$; this entails different deformations in each of the three example structures, see Fig.~\ref{fig:fig1}(b). These deformations leave the symmetry subgroup generated by $\{C_3, \sigma_{x}, T\}$ intact; the resulting space group is isomorphic to \textit{p3m}, and the corresponding symmetry-breaking perturbation takes the form~\cite{supmat}
	\begin{equation} \label{eq:M1}
	M_1 = c_1 \mathbb{1}_4 + m_1 \Gamma_1\,,
	\end{equation}
	where $\Gamma_1 = \tau_3 \otimes \tau_2$, and $c_1$ and $m_1$ are real coefficients that quantify the strength of the perturbation. Breaking translation symmetry, instead, we consider an enlarged real-space unit cell~[Fig.~\ref{fig:fig1}(c)], which leads to band folding into a smaller Brillouin zone (BZ) in reciprocal space. The larger unit cell maps both valleys $K/K'$ to the new $\Gamma$ point. Applying the appropriate deformation for each lattice then breaks the original translation invariance so that the system is no longer self-coincident under $T$. The remaining symmetry subgroup is generated by $\{C_6, \sigma_{x}, T C_6 T\}$, where $T C_6 T$ is the expanded lattice translation. The corresponding translation symmetry-breaking perturbation matrix is~\cite{supmat}
	\begin{equation} \label{eq:M2}
	M_2 = c_2 \mathbb{1}_4 + m_2\Gamma_2\,, 
	\end{equation} 
	with $\Gamma_2 = \tau_1 \otimes \mathbb{1}_2$, and real coefficients $c_2, m_2$.
	
	The perturbation matrices $\Gamma_1$ and $\Gamma_2$ anticommute. Hence, when the band gaps of the two perturbed structures coincide in energy (i.e., when $c_1 = c_2$), the identity terms in Eqs.~\eqref{eq:M1} and~\eqref{eq:M2} can be absorbed into the unperturbed Hamiltonian, leaving us with two anticommuting mass terms gapping the $K/K'$ valleys. Note that while the enlarged unit cell is not primitive in the case of inversion-breaking $M_1$, choosing such a cell allows both perturbations to be combined in a single lattice. 
	To complete the construction, we study the dispersion away from the two valleys; quasimomenta away from $K/K'$ similarly result in perturbative terms, entering Eq.~\eqref{eq:H} as~\cite{Saba_2017,Saba_2020,supmat}
	\begin{equation} \label{eq:gammas}
	\gamma_1 = \tau_3 \otimes \tau_1 \, , \qquad \gamma_2 =  \tau_3 \otimes \tau_3 \,.
	\end{equation} 
	The two matrices $\gamma_1$ and $\gamma_2$ anticommute with each other and with both $\Gamma_1$ and $\Gamma_2$. As the symmetry-breaking perturbations may be position-dependent, we have effectively arrived at the term $\boldsymbol{\Gamma} \vdot \boldsymbol{m}(\boldsymbol{r})$ and fully identified the ingredients for realizing the Jackiw-Rossi Hamiltonian in a continuous 2D medium.
	
	We highlight that our analysis is applicable to a broad range of geometries and systems, as it relies solely on the symmetries of the perturbations and on time reversal. Moreover, our construction does not rely on introducing pseudo-spin or pseudo-time reversal symmetries~\cite{Wu_Hu_2015}. Instead, the initial 4-fold degeneracy is directly obtained by starting with the original unit cell and its associated space group. We can thus also directly verify that our construction falls into the BDI class: in our chosen 4D basis, the time-reversal operator reads $\mathcal{T} = I \mathcal{K}$, where $I = \tau_1 \otimes \mathbb{1}_2$ corresponds to the spatial inversion operator and $\mathcal{K}$ is the complex conjugation operator; chiral symmetry is given by $\mathcal{C} = \tau_2 \otimes \mathbb{1}_2$. The Hamiltonian obeys these symmetries due to the spatial symmetries required by our construction. Since $\mathcal{C}^2 = \mathcal{T}^2 = 1$, our system belongs to the BDI class~\cite{Chiu_2016, Teo_2010} and thus admits a topological winding number [cf.~Eq.~\eqref{eq:invariant}].
	
	\begin{figure}
		\centering
		\hspace*{-5mm}
		\includegraphics[width=\columnwidth]{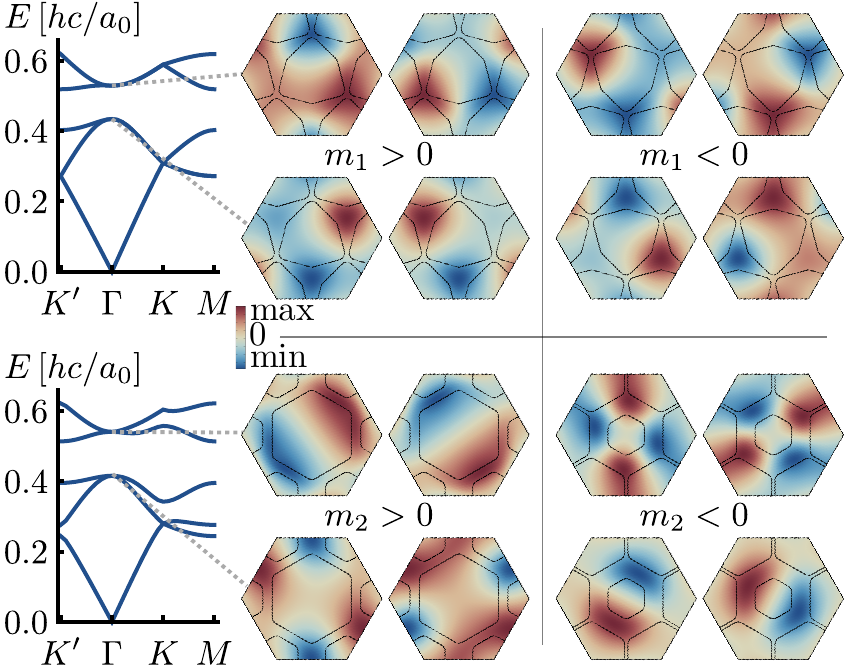}
		\caption{The bulk solutions of the perturbed simple hexagonal lattice using the enlarged unit cell with periodic boundary conditions, cf.~Fig.~\ref{fig:fig1}(I). The bulk spectrum and the four bulk eigenstates at the $\Gamma$ point below and above the gap for each perturbed unit cell structure, with positive/negative values for each of the symmetry-breaking mass terms. (top) Breaking inversion symmetry [cf.~Fig~\ref{fig:fig1}(b) and Eq.~\eqref{eq:M1}]. (bottom) Breaking translation symmetry [cf.~Fig.~\ref{fig:fig1}(c) and Eq.~\eqref{eq:M2}].}
		\label{fig:fig2}
	\end{figure}
	
	The signs of $m_1$ and $m_2$ distinguish four topologically distinct phases~\cite{supmat}. We offer an intuitive view of this distinction by examining the induced band-inversions in the expanded unit cell of the simple hexagonal lattice. We focus on the symmetries at the $\Gamma$ point and consider the point group $C_{3v}$. We also choose bases for two instances of the $C_{3v}$ twofold degenerate representation, e.g., the pairs of orbitals $\{p_x, p_y\}$ and $\{d_{xy}, d_{x^2-y^2}\}$. These can mix freely under $C_{3v}$ symmetry, and are guaranteed to form a fourfold degeneracy when the original unit cell is intact. Under the perturbation $M_1$~(Fig.~\ref{fig:fig2}, upper panel), the two distinct orientations of the enlarged unit cell are mapped into each other by inversion, fixing the eigenspaces $\{p_x + d_{xy}, \: p_y + i d_{x^2-y^2}\}$ and $\{p_x - d_{xy}, \: p_y - i d_{x^2-y^2}\}$; these match realizations of the QVHE~\cite{Ma_2016, Lu_2017}. Under the perturbation $M_2$~(Fig.~\ref{fig:fig2}, lower panel), the enlarged unit cell remains inversion-symmetric, so that its eigenspaces must consist of even and odd functions, here $\{p_x,\: p_y\}$ and $\{d_{xy},\: d_{x^2-y^2}\}$, matching the crystalline realizations of the QSHE~\cite{Wu_Hu_2015}. 
	
	\textit{Bulk signatures.---} Our discussion has so far relied on general symmetry arguments. At the same time, we show in Fig.~\ref{fig:fig1} examples of lattices and their deformations. 
	In each structure, the magnitudes of the mass terms $m_1$ and $m_2$ are evaluated numerically as they depend on the specifics of the symmetry-breaking deformations and material/platform realization. To extract the mass terms from the numerics, we focus on the $\Gamma$ point of the enlarged unit cell and obtain the four numerical solutions $H\ket{\psi_i} = E_i\ket{\psi_i}$, with energies $E_i$ and corresponding deviations from mid-gap $\Delta E_i=E_i-\bar{E}$ for $\bar{E}=\sum_i E_i/4$. In the effective $4 \times 4$ model, the mass terms are easily evaluated, since $m_i = \expval{\Gamma_i}{\psi_j}$. However, as seen in Fig.~\ref{fig:fig2}, the numerical solutions are obtained in a real-space basis. We therefore need to convert the matrices $\Gamma_1$ and $\Gamma_2$ to real space operations, i.e., invert the function $\rho(g)$ in Eq.~\eqref{eq:operations}. Averaging over the four solutions then yields
	\begin{equation}
	\label{eq:massformula}
	m_i = \sum_i \Delta E_j  \expval{\rho^{-1} (\Gamma_i) }{\psi_j}/4\,,
	\end{equation}
	where we find using Eq.~\eqref{eq:operations}
	\begin{gather}
	\mathcal{\rho}^{-1}(\Gamma_1) = -\left(4C_3 T + 2 C_3 + 2 T + \mathbb{1} \right) /\, 3\,, \nonumber\\
	\rho^{-1}(\Gamma_2) = -C_6^3\,.
	\end{gather}
	
\begin{figure}
		\centering
		\hspace*{-5mm}
		\includegraphics[width=\columnwidth]{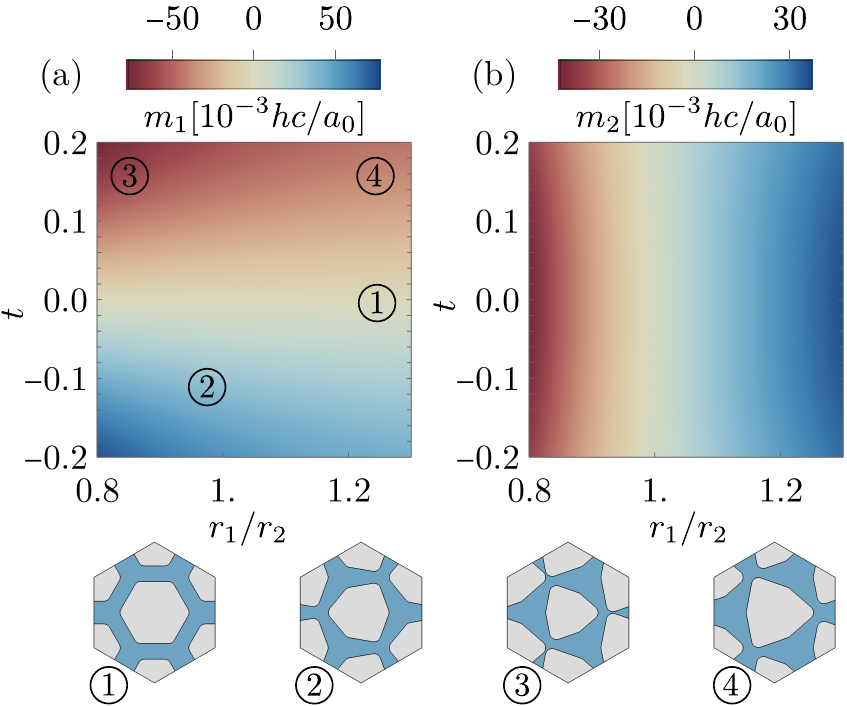}
		\caption{The bulk solutions of the perturbed simple hexagonal lattice using the enlarged unit cell and combining the symmetry-breaking deformations, cf.~Fig.~\ref{fig:fig1}(I). The (a) $m_1$ and (b) $m_2$ mass terms, extracted using~Eq.~\eqref{eq:massformula}. The bottom panel shows the unit cell at selected points in the configuration space.}
		\label{fig:fig3}
	\end{figure}

	In Fig.~\ref{fig:fig3}, we plot the dependence of the effective masses $m_i$ on the structural features of the simple hexagonal lattice example, cf.~Fig.~\ref{fig:fig1}(I). Here, inversion symmetry is broken by deforming the hexagonal incision into a triangle, parameterized by $t$. Translation symmetry is broken by introducing two different radii $r_1$ and $r_2$ for the central and outer hexagonal incisions, parameterized by $r_1/r_2$. Using such a construction, the deformations can be combined in a single unit cell, cf.~Figs.~\ref{fig:fig3}(1)-(4). The calculations are performed using COMSOL~\cite{comsol}, and the mass terms are evaluated using Eq.~\eqref{eq:massformula}. Crucially, the mass terms are varied independently by the different symmetry-breaking deformations, see Figs.~\ref{fig:fig3}(a) and (b). This confirms that we can readily realize a spatially-dependent mass term $\boldsymbol{m}(\boldsymbol{r})$, as required by the Jackiw-Rossi model. We reiterate that to create a bound mode, the structures used for the winding must have coincident band gaps, analogously to $c_1 = c_2$ in Eqs.~\eqref{eq:M1} and~\eqref{eq:M2}.
	
\begin{figure}
		\centering
		\includegraphics[width=\columnwidth]{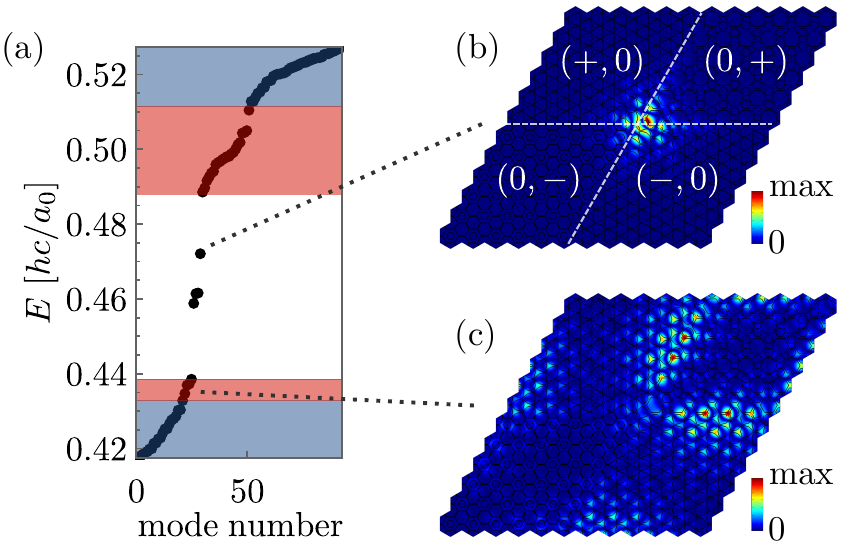}
		\caption{Numerical simulation of a Jackiw-Rossi vortex with $n = 1$ in a continuous dielectric ($\epsilon_r = 11.7$) patterned with air incisions. The simulated supercell combines four bulks corresponding to the four perturbed structures (1)-(4) in Fig.~\ref{fig:fig3}. This gives rise to a Jackiw-Rossi vortex~\eqref{eq:H} at the cell's center. Periodic boundary conditions are used at all four edges of the cell, giving a total of four vortices. (a) The energy spectrum of the structure showing bulk modes (blue fill), edge modes (red fill) and 0D bound modes (white fill). (b) The supercell, showing the signs of the mass terms $(m_1, m_2)$ and boundaries between the phases. The amplitude of a mid-gap solution is overlayed onto the structure, exhibiting a mode bound at the center. (c) A similar plot for one of the gapped edge mode solutions.}
		\label{fig:fig4}
	\end{figure}

	We move now to construct a Jackiw-Rossi vortex in a large supercell composed of the different bulk phases. To obtain a topological bound mode, the mass vector must wind [cf.~Eq.~\eqref{eq:invariant}], i.e., we need to spatially pass through all four `pure` perturbations shown in Fig.~\ref{fig:fig2}. 
	Constructing a vortex of $n=1$ in a dielectric medium, we numerically confirm the existence of a bound mode in Fig.~\ref{fig:fig4}. To assure an overall band gap, we use only the four unit cells shown in Fig.~\ref{fig:fig2}. Since we use periodic boundary conditions, the supercell features four distinct vortex sites, all with the same $\abs{n}$ but with different terminations. Correspondingly, we find four bound modes, each localized at one of the four vortices. By construction, the topological vortex modes should appear mid-gap. One of the vortex modes indeed lies at the bulk gap center, while the remaining three deviate slightly away [see Fig.~\ref{fig:fig4}(a)]. We attribute this discrepancy to the inequivalent vortex terminations, where the spatial symmetries are locally broken~\cite{Gao_2020}. Note that gapped topological edge states exist at the domain walls between different mass regions, as expected in second-order topological insulators~\cite{Benalcazar_2017a,Benalcazar_2017b,Petrides_2020}, see Figs.~\ref{fig:fig4}(a)~and~(c).
	
	\textit{Discussion.---} The presented symmetry principles and formulas for numerical calculations enable a systematic exploration of a plethora of structures, which feature high-order topological defects. The range of compatible structures is far broader than with the TB approach, e.g., the primitive cell of the simple hexagonal (Kagome) lattice has 1 (3) incisions, which precludes the formulation of a 4-band TB model required by a 2D HOTI.
	At the same time, we emphasize that the symmetry arguments describe the behavior of the bands near the high-symmetry points of the BZ, but do not guarantee a fully gapped spectrum. In the Kagome lattice, for example, a dielectric patterned with air incisions shows the correct symmetry breaking at the $\Gamma$ point, but the spectrum is gapless due to the band dispersion at other regions of the BZ, yielding only an indirect gap at the $\Gamma$ point. Such limitations are peculiar to specific experimental platforms. 
	From a technological standpoint, the bulk band gap size is the key characteristic to keeping the vortex modes spectrally isolated and spatially confined. In Fig.~\ref{fig:fig4}(a), its size relative to the gap center is $16.6\%$, which is a factor of $3$ higher compared to existing works based on the graphene-inspired TB model~\cite{Gao_2020}. Further improvement is expected with the use of numerical optimization enabled by our work. We also note that the spatial confinement may be strengthened by nonlinear interactions~\cite{Smirnova_2020, Kraus_2014}. Finally, we highlight that our scheme can be extended to three dimensions by appropriately extruding the geometry in the out-of-plane direction. This yields one-dimensional, counter-propagating topological modes that can be used to construct optical fibers~\cite{Lin_2020, Lu_2018}.

	\begin{acknowledgements}
		This work was supported by the Swiss National Science Foundation through grants CRSII5 $177198/1$ and PP00P2$\_$163818. We also thank I. Petrides, M. Rechtsman, T. Wolf, A. Eichler, and R. Chitra for helpful discussions.
	\end{acknowledgements}
	
	\bibliography{bibliography}
	
\end{document}


\title{Supplemental Material for "Second-order topological modes in two-dimensional continuous media"}
	
\author{Jan Ko\v{s}ata}
\author{Oded Zilberberg}
\affiliation{Institute for Theoretical Physics, ETH Z{\"u}rich, 8093 Z{\"u}rich, Switzerland}
\date{\today}
\maketitle

\section{Representations of symmorphic space groups} \label{sec:repretheory}

In this section, we provide a minimalist overview of the formalism needed to construct the representation matrices introduced in Eq.~(\ref*{eq:operations}) of the main text. We follow the nomenclature used in Ref.~\cite{Bradley_2009}, where our exposition is simplified by focusing on symmorphic space groups.

\subsection{Glossary}

\textit{Symmorphic space group:} The group $\mathcal{G}$ generated by a set composed of point group operations and real space translations is called symmorphic~\footnote{Equivalently, a symmorphic group has no glide planes or screw axes.}.

\textit{Isogonal group:} The point group operations of a symmorphic space group $\mathcal{G}$ form its isogonal group $\mathcal{F}$.

\textit{Equivalent $k$-points:} Two points in $k$-space are equivalent if they differ by a combination of reciprocal lattice vectors.

\textit{Little co-group:} For a space group $\mathcal{G}$, the operations which take a $k$-space vector $\boldsymbol{k}$ into an equivalent point form the little co-group of $\boldsymbol{k}$, denoted ${{\overline{\mathcal{G}}}_{\boldsymbol{k}}} \subset \mathcal{F}$.

\textit{Star:} The set of inequivalent $k$-points generated from $\boldsymbol{k}$ by all operations of a group $\mathcal{G}$ is called the star of $\boldsymbol{k}$.

\textit{Representation:} The representation $\rho$ of a group $\mathcal{G}$ is a homomorphism of $\mathcal{G}$ onto a group $\mathcal{T}$ of non-singular linear operators acting in a finite-dimensional vector space $V$.

\textit{Irreducible representation (irrep):} If $V$ has no proper subspace invariant under the operations of $\mathcal{T}$ except $V$ itself, then $\rho$ is an irrep.

\textit{Basis of an irrep:} A set of functions $B$ that spans the domain $V$ of an irrep $\rho$ is a basis of $\rho$. Each basis belongs to one irrep only.

\textit{Representation matrix:} Having chosen a basis $B$, every symmetry element $g \in \mathcal{G}$ is mapped by $\rho$ onto a matrix $\rho(g) \in \mathcal{T}$. This is the representation matrix of $g$ in the basis $B$.
\\

\subsection{Constructing an irrep and its basis}
We now consider a space group $\mathcal{G}$ with the isogonal group $\mathcal{F}$, and seek a basis $B = \{\phi_j \}$ for a yet-unknown irrep $\rho$. As we deal with periodic metamaterials, Bloch's theorem sets the form for the individual basis functions,
\begin{equation} \label{seq:blochfunc}
 \phi_j(\boldsymbol{r}) = u_j(\boldsymbol{r}) e^{i \boldsymbol{k} \vdot \boldsymbol{r}}\,,
\end{equation}
where $\boldsymbol{k}$ is a $k$-space vector, and $u_j(\boldsymbol{r})$ is invariant under lattice translations. In such a basis, the representation matrix of translation by a unit cell, $\rho(T)$, is diagonal. Choosing a vector $\boldsymbol{k}$, we now proceed in three steps.

(i)~Determine ${\overline{\mathcal{G}}}_{\boldsymbol{k}}$. For every $g \in {\overline{\mathcal{G}}}_{\boldsymbol{k}}$, we have $g \left(u_j\, e^{i \boldsymbol{k} \vdot \boldsymbol{r}}\right) = e^{i \boldsymbol{k} \vdot \boldsymbol{r}} \,g u_j$ by definition, up to a periodic function which can be absorbed into the definition of $u_j$. To be closed under ${\overline{\mathcal{G}}}_{\boldsymbol{k}}$, the set $\{u_j\}$ must therefore form a basis for its representation. Let us then choose a $b$-dimensional irrep $\rho_{\boldsymbol{k}}$ of $\overline{\mathcal{G}}_{\boldsymbol{k}}$., and a basis $B_{\rho_{\boldsymbol{k}}}$ of $\rho_{\boldsymbol{k}}$ such that every $u_j \in B_{\rho_{\boldsymbol{k}}}$ is invariant under lattice translations. Each $u_j \in B_{\rho_{\boldsymbol{k}}}$ can be used to define $u_j\, e^{i\boldsymbol{k}\vdot\boldsymbol{r}} \in B$. We have thus constructed $b$ basis functions of the irrep $\rho$.

(ii)~Take an element $h \in \mathcal{F}$ that is not in ${\overline{\mathcal{G}}}_{\boldsymbol{k}}$. As such, $h$ maps a Bloch function labeled by $\boldsymbol{k}$ into another member of its star, $h \left(e^{i \boldsymbol{k}\vdot\boldsymbol{r}}\right) = e^{i \boldsymbol{k}'\vdot\boldsymbol{r}}$. The transformed basis $h \left( B_{\rho_{\boldsymbol{k}}}\right)$ is still a basis of $\rho_{\boldsymbol{k}}$, although generally ${B_{\rho_{\boldsymbol{k}}}} \neq h \left( B_{\rho_{\boldsymbol{k}}}\right)$. From each function $h  \left(u_j\right) \in h \left( {B_{\rho_{\boldsymbol{k}}}}\right)$, we again generate a $h  \left(u_j\right)\, e^{i \boldsymbol{k}'\vdot\boldsymbol{r}} \in B$.

(iii)~Repeat step (ii) for each inequivalent $\boldsymbol{k}$ generated by $h$. 
\\

For a $\boldsymbol{k}$ with a $q$-membered star and a $b$-dimensional irrep of ${\overline{\mathcal{G}}}_{\boldsymbol{k}}$, we have thus generated a $(qb)$-dimensional basis for an irrep $\rho$ of $\mathcal{G}$. In the band diagram, this appears as $q$ degeneracy points, each $b$-fold degenerate. The representation matrix of each operation in $\mathcal{G}$ is found explicitly by its action in the basis $B$.

The application to the $K$ point of the space group \textit{p6m} is now straightforward. The isogonal point group of \textit{p6m} is $C_{6v}$. The little co-group $\overline{\mathcal{G}}_K$ is $C_{3v}$ and the star of $K$ consists of two points, $\{ K, K'\}$. Seeking a fourfold degeneracy, we focus on the 2-dimensional irrep of $C_{3v}$. A possible basis for this irrep is $(x,y)$;  we therefore define the functions $(p_x, p_y)$, which transform as $(x,y)$ under point group operations, but are invariant under lattice translations. Applying steps (i) and (ii) with $h = C_6$ gives four basis functions of \textit{p6m},
\begin{equation} \label{seq:basis}
    \{ p_x e^{i \boldsymbol{K} \vdot \boldsymbol{r}}, p_y e^{i \boldsymbol{K} \vdot \boldsymbol{r}}, p_x e^{i \boldsymbol{K'} \vdot \boldsymbol{r}}, p_y e^{i \boldsymbol{K'} \vdot \boldsymbol{r}} \} \,.
\end{equation}

Let us now find the representation matrices of the symmetry elements. The translation $T$ is diagonal, containing phase shifts due to the $K/K'$ labels,
\begin{equation}
    \rho(T) = \text{diag}\{ e^{4 \pi /3}, e^{2 \pi / 3} \} \otimes \mathbb{1}_2\,.
\end{equation}

The mirror plane $\sigma_x \notin \overline{\mathcal{G}}_K$ maps $(p_x,p_y) \rightarrow (p_x, -p_y)$ and $(K, K') \rightarrow (K', K)$, and thus gives
\begin{equation}
    \rho(\sigma_x) = \sigma_1 \otimes \sigma_3.
\end{equation}

Finally the rotation $C_6 \notin \overline{\mathcal{G}}_K$ maps $(p_x, p_y) \rightarrow R_6 (p_x,p_y)$, where $R_6$ is the $\pi/3$ rotation matrix, and $(K, K') \rightarrow (K', K)$, giving in our basis choice
\begin{equation}
    \rho(C_6) = \sigma_1 \otimes R_6\,.
\end{equation}
Note that while this proves the existence of such an irrep, the particular choice of basis is unimportant. Our numerical solutions do not immediately resemble the form constructed in Eq.~\eqref{seq:basis}. 

\section{Perturbation theory}

In this section, we apply the concepts introduced in Sec.~\ref{sec:repretheory} to obtain symmetry-allowed perturbative terms in the Hamiltonian. For an alternative discussion of the arguments we employ, see the Supplemental Material of Ref.~\cite{Saba_2017}.
We start with the eigenproblem
\begin{equation} \label{seq:eigenproblem}
    M(\boldsymbol{\lambda})  \boldsymbol{v} = E(\boldsymbol{\lambda}) \boldsymbol{v}
\end{equation}
where $M$ is a linear operator (e.g., the Hamiltonian) and $\boldsymbol{\lambda}$ is a vector of parameters. 
\\

\textit{Zeroth order:~} Suppose that the unperturbed operator $M(0)$ has a set of symmetries forming the space group $\mathcal{G}$. Take a set $V$ of vectors that correspond to a single eigenvalue $E(0)$.  If there are no accidental degeneracies, $V$ constitutes a basis for an irreducible representation of $\mathcal{G}$, thus uniquely defining the set of representation matrices $\rho(g)$. Correspondingly, for every $\boldsymbol{v} \in \text{span}(V)$ and $g \in \mathcal{G}$, we obtain a $\rho(g) \boldsymbol{v} \in \text{span}(V)$, such that
\begin{equation}
    M(0)\boldsymbol{v} = \rho(g)^{-1} M(0) \rho(g) \boldsymbol{v} =  E(0) \boldsymbol{v}\,.
\end{equation}
Since $\boldsymbol{v}$ is arbitrary, we arrive at the kernel equation
\begin{equation} \label{seq:zeroconstraint}
    M(0) = \rho(g)^{-1} M(0) \rho(g)\,,
\end{equation}
which, by construction, is satisfied only if $M(0) \propto \mathbb{1}$.
\\

\textit{First order: } We now perturb with $\boldsymbol{\lambda}$ to the first order. Keeping the same basis. Eq.~\eqref{seq:eigenproblem} now reads
\begin{equation} \label{seq:perteq}
\left[\boldsymbol{\lambda} \vdot \boldsymbol{M}_{1}(\boldsymbol{\lambda})\right] \boldsymbol{v}_i = E_1(\boldsymbol{\lambda}) \boldsymbol{v}_i\,,
\end{equation}
where we defined $\boldsymbol{M}_{1}(\boldsymbol{\lambda}) = \boldsymbol{\nabla_\lambda} M(\boldsymbol{\lambda} )$ and $E_{1}(\boldsymbol{\lambda}) = \boldsymbol{\lambda} \vdot \boldsymbol{\nabla_\lambda} E(\boldsymbol{\lambda} )$. 
For each $g \in \mathcal{G}$, transforming both sides and asserting the invariance of $E_1(\boldsymbol{\lambda})$ gives a condition analogous to Eq.~\eqref{seq:zeroconstraint}, with one important difference: the operator $\boldsymbol{\nabla}_{\boldsymbol{\lambda}}$ may itself transform non-trivially under the symmetry operations,
\begin{equation}
    \boldsymbol{\nabla}_{\boldsymbol{\lambda}} \xrightarrow[]{g} R(g)   \boldsymbol{\nabla}_{\boldsymbol{\lambda}}\,,
\end{equation}
generating the condition
\begin{equation} \label{seq:condition}
\partial_{\lambda_{j}} M(\boldsymbol{\lambda} ) = \rho(g)^{-1} R(g)_{ij} \partial_{\lambda_{i}} M(\boldsymbol{\lambda} ) \rho(g) \, .
\end{equation}
Crucially, a symmetry-allowed perturbation term $\boldsymbol{\nabla_\lambda} M(\boldsymbol{\lambda} ) $ must satisfy Eq.~\eqref{seq:condition} for each $g$ in the generating set of $\mathcal{G}$.
\\

\subsection{Quasimomentum k}

Moving $\boldsymbol{k}$ away from the high-symmetry points induces, to first order, a perturbation with $\boldsymbol{\lambda} = \boldsymbol{k}$. The operator $\boldsymbol{\nabla_{k}}$ transforms similarly to the vector $(x, y)$, for which the $2 \cross 2$ matrices $R(g)$ are readily found. Using the representation matrices $\rho(g)$ found in Sec.~\ref{sec:repretheory} and Eq.~\eqref{seq:condition} then produces a set of equations with 32 variables (one for each element of the two perturbative matrices). The solution of these equations provides the allowed perturbation terms
\begin{equation}
    k_x \left(\tau_3 \otimes \tau_1 \right)\,, \quad k_y \left(\tau_3 \otimes \tau_3 \right)\,.
\end{equation}
Note that this approach is essentially equivalent to that of $\boldsymbol{k} \vdot \boldsymbol{p}$ theory.
\\

\subsection{Structural perturbations}

If the spatial structure itself is altered, the perturbed system will generally belong to some space group $\mathcal{G'} \subset \mathcal{G}$. Only the elements $g' \in \mathcal{G'}$ now bring the lattice into self-coincidence, i.e., $R(g') = 1$, while $R(h) = 0$ for  $h \notin \mathcal{G'}$. The set of constraints [cf.~Eq.~\eqref{seq:condition}] is thus relaxed to only include the elements of $\mathcal{G'}$. Using the results of Sec.~\ref{sec:repretheory}, and solving the resulting set of equations (16 variables for each structure), we obtain the perturbation terms $\Gamma_1$ and $\Gamma_2$ shown in Eq.~(\ref*{eq:M1}) and Eq.~(\ref*{eq:M2}) of the main text.


\section{Vortex modes in the simple hexagonal lattice}
\begin{figure}
    \centering
    \includegraphics{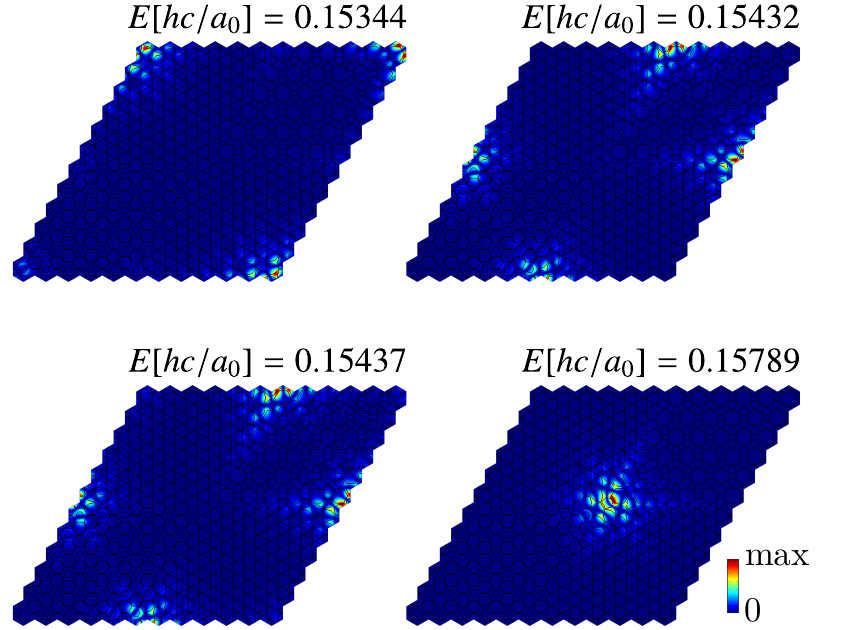}
    \caption{Numerical bound mode solutions of a Jackiw-Rossi vortex with $n = 1$ in a continuous dielectric ($\epsilon_r=11.7$) patterned with air incisions. Periodic boundary conditions are used at all four edges of the cell, giving a total of four vortices.}
    \label{sfig:modes}
\end{figure}

In Fig.~\ref{sfig:modes} we attach the plots of all four topological modes obtained in the $n=1$ supercell derived from the simple hexagonal lattice (see Fig.~\ref{fig:fig4} of the main text). Due to the finite size of the supercell, the three modes with similar energies hybridize, while the remaining mode does not.

\pagebreak
\bibliography{bibliography_supmat}